\documentstyle[preprint,epsfig,times,aps]{revtex}

\begin{document}

\draft
\tightenlines

\title{Realistic Spin Glasses Below Eight Dimensions:
A Highly Disordered View}
\author{C.M.~Newman}
\address{Courant Institute of Mathematical Sciences,
New York University, New York, NY 10012}
\author{D.L.~Stein}
\address{Departments of Physics and Mathematics, University of Arizona,
Tucson, AZ 85721}

\maketitle

\begin{abstract}

By connecting realistic spin glass models at low temperature to the highly
disordered model at zero temperature, we argue that ordinary
Edwards-Anderson spin glasses below eight dimensions have at most a single
pair of physically relevant pure states at nonzero low temperature.
Less likely scenarios that evade this conclusion are also discussed.

\end{abstract}



\section{Introduction}
\label{sec:intro}

Rigorous \cite{NS96a,NS97} and non-rigorous \cite{NS98} analyses ruling out
mean-field pictures (for an overview, see \cite{MPV,BY,Parisi00}) of
short-ranged finite-dimensional spin glass models leave open the question
of whether there is a single pair or infinitely many pure states at low
temperature (assuming that spin flip symmetry is indeed broken). The
droplet/scaling picture \cite{Mac,BM,FH86} asserts the existence of at most
a single pair in all dimensions \cite{FH87,FH88}, but the same conclusion might
arise independently of the other predictions of droplet/scaling. On the
other hand, if infinitely many pure states occur (for some dimension $d$
and temperature $T$), they should do so in accord with the chaotic pairs
picture \cite{NS97,NS92,NS96b}, in which only a single pair is seen in
typical large finite volumes, with the particular pair changing chaotically
with volume.

In this paper, we present an analysis that suggests that no multi-pair
pictures should occur below eight dimensions. The analysis is based on
relating non-zero temperature Gibbs states of Edwards-Anderson(EA) spin
glass models \cite{EA} to zero-temperature ground states of the highly
disordered model \cite{NS94,BCM}. The number of ground state pairs in the
highly disordered model is known to be one for $d < 8$ and (uncountably)
infinite for $d >8$ \cite{NS94,NS96c}. The relation is such that the number
of ground states in the highly disordered model should serve as an upper
bound for the number of (physically relevant) pure states of ordinary EA
spin glasses at low but {\it nonzero\/} temperature. We make no claim about
the number of ground states of the ordinary models at {\it zero\/}
temperature (except by other arguments for $d=2$ \cite{NS00}).

The idea behind the analysis is as follows.  We construct a one-parameter
family of EA models, parameterized by $\lambda(\beta)$, where $\beta$ is
the inverse temperature, such that $\lambda\to\infty$ (slowly) as
$\beta\to\infty$ in a way to be specified shortly.  The construction is
such that for any finite $\beta \ge \beta_0$, the model is expected to have
the same thermodynamic behavior as the ``ordinary'' EA nearest-neighbor
Ising spin glass in the same dimension corresponding to $\lambda(\beta_0)$,
in the sense that if the ordinary EA model is in a low-temperature
broken-symmetry spin glass phase at $\beta_0$, our model should be also,
and the number and organization of pure states in the two models should be
identical.  At $T=0$ (i.e., $\beta=\infty$), our model becomes the highly
disordered spin glass \cite{NS94,BCM}, whose ground state structure is
known \cite{NS94,NS96c}.  So the behaviors of the two models may diverge at
$T=0$, i.e., the ground state structure of the highly disordered models may
differ from that of the ordinary EA spin glass.  But the crucial point is
that {\it any $T>0$ Gibbs measure, restricted to any finite volume, of our
model becomes supported, as $T \to 0$, on spin configurations that are
ground states of the highly disordered model within that volume\/}.

This results in three possible conclusions, any of which is interesting.
The most reasonable follows from the natural expectation that the number of
ground states is at least as great as the number of pure states at low
temperatures; thus the number of ground states in the highly disordered
spin glass should give an upper bound to the number of low-temperature pure
states in an ordinary EA model. (There would however be no implication for
the number of {\it ground\/} states in the EA model at $T=0$.)  Because the
highly disordered spin glass has only a single pair of ground states below
eight dimensions \cite{NS94,NS96c,rigor}, the conclusion would be that the
EA spin glass has no more than a single pair of pure states for all $d<8$
at very low temperature.  
This conclusion would extend to $d=8$ if it
were shown that the highly disordered model has only a single pair of
ground states also in that dimension.

We will discuss below why the analysis and this conclusion should apply
only to {\it incongruent\/} \cite{HF87} pure states.  Two distinct pure
states that are not global flips of each other may be either incongruent or
regionally congruent.  Spin configurations chosen from incongruent pure
states have a nonvanishing density of relative domain walls, i.e. couplings
satisfied in one but not the other spin configuration; otherwise they are
regionally congruent.  If incongruence occurs, it should be generated using
different sequences of coupling-independent boundary conditions, but to see
regional congruence should require choices of boundary conditions that are
coupling-dependent.

Two other possibilities, less straightforward but also intriguing, could
conceivably occur.  One of course is that the supposition that the number
of (incongruent) ground states at $T=0$ is no smaller than the number of
(incongruent) pure states at low $T>0$ does not hold for this class of
models.  Another is that there may be no continuity or even
monotonicity at all in the {\it number\/} of pure states as a function of
temperature in the spin glass --- perhaps the most interesting but also
least likely of the three possibilities raised here \cite{note1}.

All three of these conclusions will be discussed in greater detail;
we now turn to a discussion of the construction of the class of
models to be analyzed.

\section{The model}
\label{sec:model}

We will study the Edwards-Anderson (EA) Hamiltonian \cite{EA} on ${\bf
Z}^d$, the cubic lattice in $d$ dimensions:
\begin{equation} 
\label{eq:EA}
{\cal H}_{\cal J}(\sigma)= -\sum_{\langle x,y\rangle} J_{xy} \sigma_x \sigma_y\quad ,
\end{equation}
where ${\cal J}$ denotes a realization of the couplings $J_{xy}$ and where
the brackets indicate that the sum is over nearest-neighbor pairs only,
with the sites $x,y\in {\bf Z}^d$.  We will take the spins $\sigma_{x}$ to be
Ising, i.e., $\sigma_{x}=\pm 1$.

Eq.~(\ref{eq:EA}) is the EA Ising Hamiltonian for an infinite-volume spin
glass on ${\bf Z}^d$; we also need to define the EA model on a finite
volume, given specified boundary conditions.  Let $\Lambda_L$ be a cube of
side $2L+1$ centered at the origin; i.e., $\Lambda_L= \{-L, -L+1, \cdots
,L\}^d$.  The finite-volume EA Hamiltonian is then just that of
Eq.~(\ref{eq:EA}) confined to the volume $\Lambda_L$, with the spins on the
boundary $\partial\Lambda_L$ of the cube obeying the specified boundary
condition.  (The boundary $\partial\Lambda_L$ of the volume $\Lambda_L$
consists of all sites not in $\Lambda_L$ with one nearest neighbor
belonging to $\Lambda_L$.)

The couplings $J_{xy}$ are quenched, independent, identically distributed
random variables; throughout the paper we will assume their common
distribution to be symmetric about zero.  Most studies use either the
Gaussian or $\pm J$ distributions, under the assumption that the
qualitative thermodynamic properties in fixed dimension --- existence of a
phase transition at some $T_c(d)$ (whose value, but presumably not its
existence, will depend on the nature of the distribution), the presumed
broken spin-flip symmetry (i.e., a nonzero EA order parameter $q_{EA}$)
below $T_c$, the number of pure states below $T_c$, and so on --- will be
the same for any ``reasonable'' coupling distribution.  Because this
remains an assumption, there is no precise definition of ``reasonable'',
but the expectation is that any distribution that is symmetric about zero
and falls off sufficiently quickly for larger coupling magnitudes will all
exhibit similar spin glass behavior.  Thus, for example, a uniform
distribution supported on $[-1,1]$ is expected to have the same basic
thermodynamic properties as the $\pm J$ or Gaussian spin glasses.  We
proceed using this assumption (or weakened versions of it).

We depart from previous studies in parametrizing the coupling distribution
through the variable $\lambda(\beta)$, with $\lambda\to\infty$ as
$\beta\to\infty$.  The couplings $J_{xy}$ are defined through the relation
(cf. \cite{NS94})
\begin{equation}
\label{eq:jxy}
J_{xy}^{\lambda(\beta)} = \epsilon_{xy}\ c_{\lambda(\beta)}\
e^{-\lambda(\beta)K_{xy}}
\end{equation}
where $c_{\lambda(\beta)}$ is chosen to ensure that the model has a
sensible thermodynamic limit (i.e., a finite energy/spin) when
$\beta \to \infty$, and $\epsilon_{xy}$ and $K_{xy}$ are two sets of
independent, identically distributed (i.i.d.)  random variables.  Each
variable $\epsilon_{xy}$ takes on one of the two values $\pm 1$ with equal
probability, and the $K_{xy}$ can be chosen from any continuous
distribution (e.g. uniformly from $[0,1]$) so that the distribution of the
$J_{xy}$'s is ``reasonable'', as discussed above.  Two examples will be
given shortly.

Our general approach is to start with a ``typical'' EA model at
$\lambda(\beta_0)=1$, and then embed this within a one-parameter family of
models that has a known ground state structure (corresponding to
$\lambda(\infty)=\infty$).  Let us denote by $J_0$ a given random coupling
at the origin connecting to a specified nearest-neighbor site.  By
Eq.~(\ref{eq:jxy}), for a given realization of the random variables
$\epsilon$ and $K$, we have $J_0=\epsilon_0 |J_0|=\epsilon_0\ c_1\ e^{-K_0}$
($c_1$ can be taken equal to $1$).
After a rescaling of the
temperature, we have
\begin{equation}
\label{eq:iterate}
J_0^\lambda=\epsilon_0\ c_\lambda\ |J_0^\lambda| = \epsilon_0\ c_\lambda\ 
e^{-\lambda K_0} \ .
\end{equation}
We will see that a choice of $c_\lambda \sim 1/\overline{|J_0|^\lambda}$, where
an overbar denotes an average over coupling realizations, will ensure a
sensible thermodynamics for the model in the limit of zero temperature
(providing $\lambda$ increases slowly as $\beta \to \infty$).

Note that when $\lambda=0$ the model becomes the $\pm J$ model for any
starting choice of coupling distribution.  Of more relevance is the
opposite limit: as $\lambda \to \infty$, all initial coupling distributions
merge into the highly disordered model discussed in \cite{NS94,BCM,NS96c}.
In this limit the coupling distribution is infinitely ``stretched''
nonlinearly, so
that every coupling magnitude occurs on its own scale.  A full analysis of
the ground state structure of this model, and the transition in ground
state pair multiplicity, is given in \cite{NS94,NS96c}.  A quantum version
of the highly disordered limit has been used to study the random quantum
Ising model in a transverse field \cite{MMH}.

So given a wide range of choices for the distribution of the $K$'s, the
spin glass model described here has a ``reasonable '' coupling distribution
for any $\beta<\infty$; we will in fact work with coupling distributions
that have a large-magnitude cutoff at finite values (depending on $\beta$),
which guarantees this.  As $\beta$ increases, the distribution becomes
increasingly stretched, but retains its finite large-magnitude cutoff for
any $\beta<\infty$, so that any of these models retains the qualitative
thermodynamic behavior of the ordinary EA model.

As we will see, the dependence of $\lambda$ on $\beta$ can be chosen (with
$\lambda$ increasing sufficiently slowly) so that if an ordinary EA model,
corresponding to some finite $\lambda$ (e.g., $\lambda(\beta_0)=1$), is in
a low-temperature spin glass phase, then so is the model described above
with running $\lambda$.  We will further see that if $\lambda$ increases
slowly enough with $\beta$ (depending on the choice of distribution for
$K$), then the Gibbs measures in the limit $\beta \to \infty$ are supported
only on spin configurations that are ground states of the highly disordered
model.  That is natural since taking $\lambda(\beta)\to\infty$ slowly
enough is roughly equivalent to first taking the $\beta\to\infty$ ($T \to
0$) limit and then the $\lambda\to\infty$ limit.

\subsection{Two examples}
\label{subsec:examples}

In this section we give two examples of coupling realizations that can be
used in subsequent analyses.  As already noted, the precise form of the
distribution is unimportant as long as the requirements listed earlier are
met; all such models should exhibit the same positive-temperature behavior
and will merge at zero temperature.  The first example starts with a flat
distribution for $J_{xy}$ in the interval $[-1,1]$ at $\lambda(\beta_0)=1$.
We need to determine the distribution for $K$ necessary to recover this
flat distribution for $J$.  Let
\begin{equation}
\label{eq:u}
{\cal U}^\lambda=e^{-\lambda K}
\end{equation}
with ${\cal U}^1=e^{-K}$ having a flat distribution in $[0,1]$.

It follows that $K$ is taken from an exponential distribution 
on $(0,\infty)$, with $P(K)\ dK=e^{-K}\ dK$.  Now let
$J_{xy}^\lambda=\epsilon_{xy}\ c_\lambda\ {\cal U}_{xy}^\lambda$; then
$M = |J_{xy}^\lambda|$ has the distribution
\begin{equation}
\label{eq:jflat}
P(M)\ dM =\cases{\lambda^{-1}(M/c_\lambda)^{1/\lambda}M^{-1}\ dM,
&if $0 < M < c_\lambda$\, ,\cr 
0, &otherwise\, . \cr}
\end{equation} 
It remains to choose $c_\lambda$.  The procedure for doing this
will be discussed in Sec.~\ref{subsec:energy}; we here give the result,
which is $c_\lambda = \lambda$, so that
$J_{xy}^{\lambda} = \lambda\,\epsilon_{xy}\ {\cal U}_{xy}^\lambda$
and
\begin{equation}
\label{eq:jflatdist}
P(M)\ dM = \cases{(1/\lambda)\left[M/\lambda\right]^{1/\lambda}
M^{-1} \ dM,&if
$0 < M < \lambda$\, ,\cr 
0, &otherwise\, . \cr}
\end{equation}

The second example starts with a uniform distribution for
$K$ on $[0,1]$.
Again
\begin{equation}
\label{eq:kflat}
J_{xy}^{\lambda} =\lambda\ \epsilon_{xy}\ e^{-\lambda K_{xy}}\ ,
\end{equation}
where we have already chosen $c_\lambda = \lambda$, but the
distribution of $M = |J_{xy}^{\lambda}|$ is now given by
\begin{equation}
\label{eq:kflatdist}
P(M)\ dM\,=\cases{(1/\lambda)M^{-1}\ dM,&if
$\lambda e^{-\lambda} < M\,<\lambda$\, ,\cr 
0, &otherwise\, . \cr}
\end{equation} 
The coupling distribution for this case is graphed in Fig.~1. 

\begin{figure}
\centerline{\epsfig{file=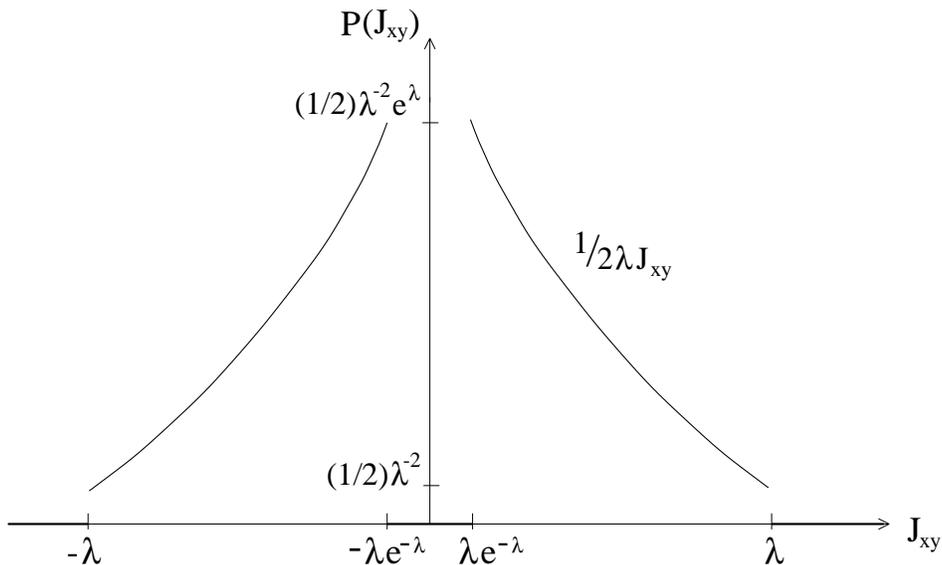,width=5.0in}}
\vspace{0.3in}
\caption{Sketch of the coupling density $P(J_{xy})\ dJ_{xy}$ from
Eq.~(\ref{eq:kflatdist}) for fixed $\lambda<\infty$, corresponding
to a uniform distribution of $K_{xy}$ on $[0,1]$.}
\label{fig:jxy}
\end{figure}

It is important to note that as $\lambda \to \infty$, the distributions of
$J_{xy}^{\lambda}$ in the two examples approach each other; this is true in
general.  Both of these distributions have a temperature-dependent cutoff
at large coupling magnitudes.  While distributions with cutoffs are chosen
for convenience, one could also choose a Gaussian distribution for $J_{xy}$
at $\lambda(\beta_0)=1$.  Because all of these distributions are expected
to give the same thermodynamic behavior, however, we will henceforth use
the simplest of these, given by Eqs.~(\ref{eq:kflat}) and
(\ref{eq:kflatdist}).

\section{Pure state structure within the metastate}
\label{sec:metastate}

We will study this class of models and will show that (for slow
enough increase of $\lambda$ with $\beta$), as $\beta \to \infty$, the
Gibbs measures (restricted to {\it arbitrary\/} fixed volumes $\Lambda_L$)
become supported on the ground states of the highly disordered spin glass.
We require that $c_\lambda$ be chosen so that the model will be sensible in
the zero-temperature limit, in that that the energy density converges to a
finite nonzero value, and further that the choice of $\lambda(\beta)$ be
made in such a way that, for each $\beta > \beta_0$, the model always
remains within the low temperature broken-spin-flip-symmetric spin glass
phase (assuming the corresponding ordinary EA model at $\beta_0$ is in such
a phase).

If these requirements can be met, then the most natural conclusion is that
in any dimension $d$, the number of ground states of the highly disordered
model provides an upper bound to the number of pure states observed at low
(but nonzero) temperature in the ordinary EA spin glass.  (We emphasize
again, though, that other conclusions remain possible, though perhaps less
plausible; these will be discussed in Sec.~\ref{sec:discussion}.)

As we have emphasized in earlier papers, if there are multiple pure states,
the interesting, and physically relevant, situation is the occurrence of
incongruent states.  Regional congruence is of mathematical interest, but
to see it would require a choice of boundary conditions carefully
conditioned on the coupling realization ${\cal J}$.  It is not currently
known how to choose such b.c.'s in spin glasses.  Numerical treatments that
look for multiple pure states implicitly search for incongruent ones.  (It
is interesting to note that recent numerical studies \cite{KM,PY00,KPY},
some employing coupling-dependent {\it bulk\/} terms in the Hamiltonian,
have suggested the possibility of observation of regionally congruent
states; but see also \cite{MP00} for a different interpretation of the
numerical data.  However, recent work indicates that the energetics of the
interfaces found in these studies are inconsistent with regionally
congruent pure or ground states \cite{NScomment}.)

We expect our analysis to hold for the number of (incongruent) pure states
within any of the coupling-independent boundary condition {\it
metastates\/}.  The concept of metastate was introduced and discussed in
Refs.~\cite{NS97} and \cite{NS96b}, and shown to be equivalent to an
earlier (but somewhat different) construct in Ref.~\cite{AW}.

A metastate is a measure on (infinite-volume) Gibbs states at fixed
temperature that is constructed via an infinite sequence of volumes
$\Lambda_L$, with specified boundary conditions on each $\partial\Lambda_L$
chosen in a coupling-independent manner.  Roughly speaking, the metastate
provides the probability (for varying large $L$) of various Gibbs states
appearing within $\Lambda_L$.  The equivalence of metastates constructed
with certain different coupling-independent boundary conditions was shown
in Ref.~\cite{NS98}.

If there are infinitely many (incongruent) pure states, a metastate should
be dispersed over them, giving their relative likelihood of appearance in
typical large volumes.  If there is no incongruence, the metastate should
be unique and supported on a single pure state pair, and that pair will
appear in most (i.e., a fraction one) of the $\Lambda_L$'s.  Regionally
congruent states, if present, would be ``invisible'' in the metastate,
i.e., would appear in a vanishing fraction of all the $\Lambda_L$'s.
Hence, even if regional congruence existed, there would still be two
special or ``preferred'' pure states that would be seen in a typical
$\Lambda_L$ with $L$ large.

An example of regional congruence is afforded by interface states in
ferromagnets, which can only be seen with carefully chosen boundary
conditions, such as those of Dobrushin \cite{Dob}.  These are b.c.'s in
which the boundary spins above the ``equator'' (a plane or hyperplane
parallel to two opposing faces of $\Lambda_L$ and cutting it essentially in
half by passing at a finite height above or below the origin) are chosen to
be plus and the boundary spins below the equator are minus.  Here the
special pair of states consists of the uniformly magnetized up and down
states, which are seen in large volumes with random boundary conditions as
well, of course, as with periodic or free b.c.'s.  (They would also be seen
even with antiperiodic b.c.'s, since the interface would pass through any
fixed finite region only in a vanishing fraction of volumes.)  An important
difference with the spin glass case is that in the latter there is no known
procedure for obtaining boundary conditions that would ``see'' any
analogous regionally congruent states.

In the current context, one needs to specify which Gibbs measures are being
examined at $T > 0$.  The procedure in which one chooses
coupling-independent boundary conditions and then varies the temperature is
well-defined when one is studying properties of the metastate as a whole.
As noted, this will provide information on the number of incongruent pure
states (in the metastate) at a given temperature.  If one wants to push
further in order to study the possible existence of regionally congruent
pure states, then one needs to pick out these Gibbs states through a
choice of coupling-{\it dependent\/} boundary conditions.  In order to take
the $\beta\to\infty$ limit, one would then have to change these
coupling-dependent boundary conditions in some unknown way; in this case,
the procedure of taking the zero-temperature limit is a priori not
well-defined in general.  For spin glasses, the only well-defined
procedures known at this time use coupling-independent b.c.'s, as $T \to
0$.

The consequence is that this argument is best able to provide information
on the number of incongruent states in the spin glass metastate at various
positive temperatures, but that it is harder to draw conclusions on the
possible existence or nonexistence of regionally congruent states.  These
latter are the interface states that are ``invisible'' in
coupling-independent metastates, and in any case, as we have argued in
earlier papers \cite{NS97,NS98,NS00}, are unlikely to be physically
relevant even if they exist, since boundary conditions used in (either
numerical simulations or) laboratory experiments on spin glasses are
coupling-independent --- i.e., they are not tailored to the microscopic
disorder configuration.  (For other results and discussion concerning the
distinction between coupling-dependent and independent boundary conditions,
see~\cite{EF,GNS93}.)

\section{Analysis}
\label{sec:analysis}

We now analyze the behavior of our class of models both at large finite
$\lambda$ (i.e., at low temperature) and in the $\lambda\to\infty$ limit.
As noted above, for specificity we will study the behavior of the model
with coupling distribution given by Eqs.~(\ref{eq:kflat}) and
(\ref{eq:kflatdist}).

Our argument relies on answering three questions (in the affirmative).  For
a given model (i.e., specified coupling distribution) can one choose a
$c_\lambda$ and a $\lambda(\beta)$ (independently of particular coupling
realization) so that:

\noindent 1) the model has a thermodynamically sensible scaling limit as
$\beta\to\infty$, in that the energy per spin
converges to some $C$ as $\beta\to\infty$, with $0<C<\infty$;

\noindent 2) $\lambda(\beta)$ scales in such a way to ensure that, if
the model is in its low temperature 
thermodynamic phase originally, i.e., at
$\lambda(\beta_0)=1$, the model at rescaled (lower) temperature remains in
that same phase;

\noindent 3) $\lambda(\beta)$ increases slowly enough 
with $\beta$ to ensure that any Gibbs
state at positive temperature is increasingly supported, as temperature is
lowered, on spin configurations that are ground states of the highly
disordered model?

In the following three subsections, we provide an analysis that answers
these questions in the affirmative, and shows how such a $c_\lambda$ and
$\lambda(\beta)$ can be chosen.

\subsection{Effective temperature and couplings}
\label{subsec:efftemp}

In this subsection we present a simple transformation
on the (inverse) temperature and the couplings that
enables us to map the class of models under study onto
more familiar ones.  In the absence of a scaling factor
$c_\lambda$, the Gibbs weighting factor at
inverse temperature $\beta'$ is
\begin{equation}
\label{eq:weight}
\exp\left[-\beta'\sum_{\langle xy\rangle}
\ \epsilon_{xy}\ e^{-\lambda K_{xy}}\sigma_x\sigma_y\right]\, .
\end{equation}
We transform the temperature and couplings using
\begin{equation}
\label{eq:transform}
\beta'\ \epsilon_{xy}\ e^{-\lambda(\beta)K_{xy}} =
\beta \ J_{xy}^{\lambda(\beta)}
\end{equation}
where $\beta' = \beta c_{\lambda(\beta)}$ and 
\begin{equation}
\label{eq:couplings}
 J_{xy}^{\lambda(\beta)} =
\epsilon_{xy}\ c_{\lambda(\beta)}\ e^{-\lambda(\beta)\,K_{xy}}\, ,
\end{equation}
so at a given $\beta'$ the model maps onto an equivalent one
at effective temperature $\beta$ and with effective 
couplings $J_{xy}^{\lambda(\beta)}$.  Although simple, this rewriting
of the Gibbs factor provides a natural separation of the various
factors that allows us to choose the prefactor $c_{\lambda}$
in a coupling-independent way, so that the energy density 
has a sensible
$\beta\to\infty$ limit.  

To achieve the correct scaling limit, we need to choose
$c_{\lambda(\beta)}$ so that it scales (with $\beta$ or $\beta'$) 
as the inverse of (minus) the
energy density $e(\beta',\lambda)$ of the model with 
Gibbs factor (\ref{eq:weight}).
This energy density $e(\beta',\lambda)$ is the same for all
(infinite-volume) Gibbs states and for almost all realizations of the
couplings (under the assumption that $\overline{|J_{xy}^{\lambda}|} <
\infty$) and so equals its disorder average $\overline{e(\beta',\lambda)}$.
Since $\beta'$ (and $\beta$) will be chosen to scale to infinity rapidly as
functions of $\lambda$, we will choose $c_\lambda$ to scale like the
inverse of (minus) $\overline{e(\infty,\lambda)}$, the ground state energy
density, which in turn scales like (minus) $\overline{|J_{xy}^{\lambda}|}$
as $\lambda \to \infty$. To justify this choice of $c_\lambda$, we need to
compute $e(\beta',\lambda)$ in the low-temperature limit.  This is done in
the next section.

\subsection{Thermodynamic behavior of the zero temperature limit}
\label{subsec:energy}

In this subsection we estimate the disorder-averaged energy per spin
$\overline{e(\beta',\lambda)}$.  Although $\beta'$ and $\lambda$ will later
be taken to depend on each other, for the purposes of this section we treat
them as independent variables.  One of the purposes of this calculation is
to provide information on how slowly $\lambda$ needs to vary with $\beta'$
(and hence with $\beta$) in order to have sensible thermodynamics in the
zero temperature limit.  In Sec.~\ref{subsec:gs} we will study how
$\lambda$ needs to vary with $\beta$ in order for pure states at positive
temperature to be supported on ground states of the highly disordered
model.  We will see that if $\lambda$ grows slowly with $\beta'$ and
$\beta$, these two calculations result in mutually compatible ranges for
allowed scaling behaviors.

We now proceed to show that the energy per spin, at large $\beta$, can be
computed as a disorder average over a single (arbitrary) coupling 
$J_0^\lambda = e^{-\lambda\,K_0}$,
which as before denotes the coupling (in ${\cal J}$) connecting the
spin at the origin with one of its nearest neighbors.  Let
$P_{\beta',\lambda}$ denote the probability (within a Gibbs measure at
fixed $\beta'$).  Then, in dimension $d$,
\begin{equation}
\label{eq:erg}
-\overline{e(\beta',\lambda)}=\,d\, \overline{\left(P_{\beta',\lambda}
(J_0^\lambda \, \hbox{ is satisfied}) -
P_{\beta',\lambda}(J_0^\lambda \, \hbox{ is unsatisfied})\right)
|J_0^\lambda |}\, .
\end{equation}

At large $\lambda$ (and larger $\beta'$), the 
main contribution to the energy density arises
from couplings corresponding to the smallest $K_{xy}$'s, and their probability
of being satisfied approaches one.  Moreover,  
\begin{equation}
\label{eq:inequality}
0\, = -\overline{e(0,\lambda)} \, \le  -\overline{e(\beta',\lambda)}
\, \le  -\overline{e(\infty,\lambda)}\, < d\overline{\ |J_0^\lambda|}\ ,
\end{equation}
where the last inequality is 
because not every coupling is satisfied at zero temperature.
So it will be sufficient for our purposes to derive a lower
bound on $-\overline{e(\beta',\lambda)}$ that approaches 
$d\overline{\ |J_0^\lambda|}$ as $\beta \to \infty$.
In arriving at a lower bound,
the following general inequality will be useful:
\begin{eqnarray}
\label{eq:sat}
P_{\beta',\lambda}(J_0^\lambda\, \hbox{ is satisfied}) -
P_{\beta',\lambda}(J_0^\lambda\, \hbox{ is unsatisfied})\nonumber\\
& = 1-2P_{\beta',\lambda}(J_0^\lambda\, \hbox{ is unsatisfied})\nonumber\\
& \ge 1-2{P_{\beta',\lambda}(J_0^\lambda\, \hbox{ is unsatisfied})\over
P_{\beta',\lambda}(J_0^\lambda\, \hbox{ is satisfied})}\ .
\end{eqnarray}

We use the coupling magnitude distribution Eq.~(\ref{eq:kflatdist}), which
arises from the flat distribution for $K$ on the unit interval $[0,1]$.
Because most of the contribution to the energy density comes from couplings
corresponding to small $K_{xy}$'s, we break up the calculation into three
cases according to the realization of $K_0$ and its $2(2d-1)$ adjacent
couplings:

\noindent Case I:  $K_0\ge\delta$,
where $\delta$ is fixed and $0<\delta\ll 1$;

\noindent Case II: $K_0<\delta$, and
one or more adjacent couplings have magnitudes
corresponding to $K\le b_0\delta$, where $b_0$ is a constant with
$1/\delta > b_0>1$;

\noindent Case III: $K_0<\delta$,
and all adjacent couplings have magnitudes
corresponding to $K> b_0\delta$.

If we denote by $e_I$ the contribution to (minus $1/d$ times)
the energy density from 
coupling realizations corresponding to Case I, and similarly for the
other cases, then trivially 

\begin{equation}
\label{eq:sum}
- (1/d) \overline{e(\beta',\lambda)}\,= \, e_I+e_{II}+e_{III}\, .
\end{equation}

We now study the three cases separately, denoting by ${\cal I}_I$,
${\cal I}_{II}$ and ${\cal I}_{III}$ the indicator functions
that equal $1$ (otherwise $0$) only for those
coupling realizations satisfying respectively the requirements
of the three different cases.

\noindent Case I:  Here we have
\begin{equation}
\label{eq:eI}
e_I \le \overline{ {\cal I}_I |J_0^\lambda|} \,
=\int_\delta^1\ e^{-\lambda K}\ dK
 =O\left((1/\lambda)\ e^{-\lambda\delta}\right)
\end{equation}
as $\lambda\to\infty$.

\noindent Case II:  The calculation here is similar; we have
\begin{equation}
\label{eq:eII}
e_{II} \le \overline{ {\cal I}_{II} |J_0^\lambda|} \,
\le 2(2d-1)b_0 \delta \int_0^\delta\ e^{-\lambda K}\ dK
= O(\delta)\int_0^\delta\ e^{-\lambda K}\ dK
\end{equation}
as $\lambda\to\infty$.

\noindent Case III: This case is more involved. There are four possible
configurations for the two spins coupled through $J_0^\lambda$; two of
these correspond to $J_0^\lambda$ satisfied and two to $J_0^\lambda$
unsatisfied.  Consider the ratio ${P_{\beta',\lambda}(J_0^\lambda\, \hbox{
is unsatisfied})\over P_{\beta',\lambda}(J_0^\lambda\, \hbox{ is
satisfied})}$ that appears in Eq.~(\ref{eq:sat}).  This ratio is maximized
by the following ``worst case'' scenario: in changing a satisfied
configuration (for $J_0^\lambda$) to an unsatisfied one, all of the $2d-1$
adjacent couplings touching the flipped spin change from unsatisfied to
satisfied.  This case maximizes the cost in the Gibbs factor for the change
in spin configurations.  It follows that under the requirements of case
III,
\begin{equation}
\label{eq:cost}
{P_{\beta',\lambda}(J_0^\lambda\, \hbox{ is unsatisfied})\over
P_{\beta',\lambda}(J_0^\lambda\, \hbox{ is satisfied})}
\le \exp[-2\beta' e^{-\lambda\delta+(2d-1)2\beta' e^{-b_0\lambda\delta}}]\ .
\end{equation}

Therefore
\begin{eqnarray}
\label{eq:eIII}
e_{III}  = & \overline{{\cal I}_{III} |J_0^\lambda| (1 - 
2 \, P_{\beta',\lambda}(J_0^\lambda\, \hbox{ is unsatisfied}))}\nonumber\\
 = & [1-O(\delta)]\left(1-O[e^{-2\beta' e^{-\lambda\delta}+
(2d-1)2\beta' e^{-b_0\lambda\delta}}]\right)\int_0^\delta\ e^{-\lambda K}\ dK 
\end{eqnarray}
as $\lambda\to\infty$.  

The behavior of the factor $e^{-2\beta' e^{-\lambda\delta}+ (2d-1)2\beta'
e^{-b_0\lambda\delta}}$ that appears in Eq.~(\ref{eq:eIII}) as
$\beta'\to\infty$ and $\lambda\to\infty$ is sensitive to the dependence of
these two parameters on each other.  We have
\begin{equation}
\label{eq:factor}
e^{-2\beta' e^{-\lambda\delta}+ (2d-1)2\beta'
e^{-b_0\lambda\delta}}=e^{-2\beta' e^{-\lambda\delta}(1-(2d-1)
e^{-(b_0-1)\lambda\delta})}\sim e^{-2\beta' e^{-\lambda\delta}}\, \hbox{ as }\,
\lambda\to\infty
\end{equation}
regardless of the detailed behavior of $\lambda(\beta)$.  Furthermore, 
$\beta' e^{-\lambda\delta}\to\infty$  for any $\delta$,
if $\lambda\to\infty$ slower
than $\log(\beta')$ (i.e., if $\lambda/{\log(\beta')} \to 0$).  
We will therefore require (for this particular coupling
distribution, but the calculation is similar for others) that
\begin{equation}
\label{eq:lambda}
\lambda=o\left[\log(\beta)\right]\, \hbox{ as } \beta\to\infty\, ,
\end{equation}
which, as we shall see, will also guarantee that 
$\lambda = o\left[\log(\beta')\right]$.

Returning to the comparison of the energy densities (as $\beta'\to\infty$)
for the three cases, we see that $e_I$ is reduced from $e_{III}$
by a factor of order $e^{-\lambda\delta}$, and $e_{II}$ is reduced from
$e_{III}$ by a factor of order $\delta$.  We therefore find that
\begin{equation}
\label{eq:edens}
-\overline{e(\beta',\lambda)}=d[1\pm O(\delta)]\int_0^1 e^{-\lambda K}\ dK\ ,
\end{equation}
so that in the joint limit $\beta' \to\infty$, $\lambda\to\infty$ (and with
the condition $\lambda = o\left[\log(\beta')\right]$
satisfied for $K$ uniformly distributed on
$[0,1]$), since $\delta$ can be chosen arbitrarily small, it follows that
\begin{equation}
\label{eq:finally}
-\overline{e(\beta',\lambda)}\, / \, (d\, \overline{|J_0^\lambda|})\to 1\, 
\hbox{ as } \beta'\to\infty\, .
\end{equation}

Therefore, if $K\in[0,1]$ uniformly, we have
\begin{equation}
\label{eq:kunif}
-\overline{e(\beta',\lambda)}\sim \overline{|J_0^\lambda|}
=\overline{e^{-\lambda K_0}} =  \int_0^1
 e^{-\lambda K}\ dK \sim 1/\lambda\, \hbox{ as }\lambda\to\infty\ ,
\end{equation}
and we set $c_\lambda=\lambda$ for this distribution, as in
Eq.~(\ref{eq:kflat}).  Thus $\beta' = \beta c_\lambda =\beta \lambda$
and (\ref{eq:lambda}) will indeed guarantee 
$\lambda = o\left[\log(\beta')\right]$.
A similar calculation for the distribution of
Eq.~(\ref{eq:jflat}), discussed in Sec.~\ref{subsec:examples}, results in
$c_\lambda \sim \lambda+1$, and so we may choose 
$c_\lambda = \lambda$.  More generally, one chooses
$c_\lambda \sim 1/\overline{|J_0^\lambda|}$, 
as discussed in Sec.~\ref{sec:model}, 
to obtain a finite, nonzero energy density as temperature goes to zero.

\subsection{Comparison to phase behavior of ordinary EA models}
\label{subsec:comparison}

We have shown that the first of our requirements, that the thermodynamics
of our class of models behave properly in the zero temperature limit, can
be met, and have shown how to compute the energy density at low
temperatures.  We turn now to our second requirement, namely that the model
remain in the low temperature spin glass phase as $T$ is lowered from a
starting temperature already within the spin glass phase for the
$\lambda(\beta_0)$ model.  Of course, the low-temperature behavior of the
ordinary EA spin glass in dimensions three and higher is not well
understood, and it could conceivably be the case that in some or all
dimensions between three and eight it undergoes a succession of phase
transitions (either at discrete temperatures or continuously), or has no
phase transition at positive temperature at all (i.e., remains paramagnetic
down to zero temperature), or has other, perhaps more exotic, behavior.
Our only goal here is to show that our class of models behaves similarly to
the ordinary EA spin glass at least for very low, nonzero temperatures. To
keep matters simple, we'll assume that for some range of $d$, the ordinary
EA spin glass undergoes a phase transition at some $T_c(d)>0$, such that
below that temperature there is a spin glass phase with broken spin flip
symmetry, and with the number (and organization) of pure states not
depending on $T$, for $0 < T < T_c(d)$.

That our class of models should behave similarly (at $\beta < \infty$) to
ordinary EA models, in terms of numbers and organization of pure states as
$\beta$ changes, is not immediately obvious because the couplings in our
models are temperature-dependent.  It does seem reasonable to expect though
that, so long as the couplings depend weakly on $\beta$, our models should
behave similarly to more conventional ones.  However, we can improve on
this expectation and show that in fact this follows from a natural
universality hypothesis.

The universality we have in mind is that the above assumption about
ordinary EA spin glasses is valid separately for each fixed finite
$\lambda$. Thus there will be a critical inverse temperature
$\beta_c(\lambda)<\infty$ for the Hamiltonian Eq.~(\ref{eq:EA}) with
couplings $J_{xy}=\epsilon_{xy}\ c_\lambda e^{-\lambda K_{xy}}$.  In order
to satisfy our second requirement, $\lambda$ has to grow slowly enough with
$\beta$ (or equivalently, $\beta$ has to grow rapidly enough with
$\lambda$) so that for $\beta \ge \beta_0$,
\begin{equation}
\label{eq:betavary}
\beta(\lambda)>\beta_c(\lambda)\, .
\end{equation}

So, for example, if $\lambda(\beta_0)=1$ and the corresponding spin glass
model is in its low $T$ spin glass phase, the inequality
(\ref{eq:betavary}) will guarantee that the sequence of models with running
$\lambda>1$ remain within their corresponding low $T$ spin glass phase.
The above inequality can always be satisfied consistently with the other
conditions 1) and 3); all that needs to be done is to choose $\lambda$ to
grow sufficiently slowly with $\beta$.  The remaining question will be
whether the constraints imposed by conditions 1) and 3) are compatible; we
reserve that for the following section.

\subsection{Ground state behavior of zero temperature limit}
\label{subsec:gs}

We now turn to a central question in the analysis of our models, which is
how slowly $\lambda$ must vary with $\beta$ in order that condition 3)
above is satisfied, i.e., whether the Gibbs states of our models at
positive temperature are increasingly supported, as temperature is lowered,
on spin configurations that are ground states of the highly disordered
model, and whether this constraint is compatible with condition 1).

Consider again a fixed (arbitrary) volume $\Lambda_L$ centered at the
origin, with {\it any\/} fixed boundary condition chosen independently of the
coupling realization.  Let $\Delta E(\lambda)$ denote the energy difference
between the lowest-energy state in $\Lambda_L$ at given $\lambda$ and the
first excited state.  Then condition 3) is satisfied if 
$\beta\Delta E(\lambda (\beta))\to\infty$
as $\beta\to\infty$, because the lowest-energy state goes
to the ground state of the highly disordered model 
(for that b.c., in that volume) as $\lambda\to\infty$.

We again study this question in the context of a particular model, namely
where $K$ is chosen uniformly from the interval $[0,1]$.  We have here that
$J_{xy}^\lambda\sim\lambda\ \epsilon_{xy}\ e^{-\lambda K_{xy}}$.  As $\lambda$
gets large for fixed $L$, the first excited state corresponds to the
(multi-)spin flip that changes only the smallest 
magnitude coupling in the (wired b.c.) invasion tree
(see Refs.~\cite{NS94,NS96c}) 
from satisfied to unsatisfied.  But
for the distribution chosen, the magnitude of this coupling is larger than
$\lambda\,\exp[-\lambda]$, so that at any $\lambda$, $\Delta E(\lambda) >
\lambda \,e^{-\lambda}$.  
Therefore,
\begin{equation}
\label{eq:excited}
\beta\Delta E(\lambda)\ge \beta\lambda e^{-\lambda}\, ,
\end{equation}
so if $\lambda$ grows as $\log(\beta)$ or slower, condition 3) is satisfied
(for this particular distribution of $K$).

Condition 1) requires a slightly stronger constraint, given by
(\ref{eq:lambda}), which is certainly compatible with the constraint given
above.  It is gratifying that both conditions require $\lambda$ to grow
slowly with $\beta$, as initially anticipated.  It might be that condition
2) would require $\lambda$ to grow even more slowly with $\beta$, but in
any event poses no conflicts with the other constraints.

The procedure given here can be adapted for other distributions, but it is
already sufficient that an explicit example can be constructed of a model
with all of the desired properties listed earlier.

\section{Disordered Ferromagnets}
\label{sec:ferro}

The basic argument of this paper is that for the models with temperature
dependent couplings as in Eq.~(\ref{eq:jxy}) (and with properly
chosen $c_\lambda$ and $\lambda(\beta)$), the absence of more than
a single pair of ground states at $\beta = \infty$ for $d < 8$
is evidence for no more than two pure states in ordinary EA
spin glass models with fixed couplings at very low temperatures.

A potential flaw in this line of reasoning seems to arise in the case of
disordered ferromagnets, corresponding to the elimination of the random
signs $\epsilon_{xy}$ in Eq.~(\ref{eq:jxy}). After all, highly disordered
ferromagnets also have only a single pair of ground states for $d < 8$
\cite{NS94,NS96c}, but ordinary disordered ferromagnets are expected to
have multiple pure states for $d \ge 4$ \cite{roughening}
--- namely, the interface states \cite{HH85,K85} obtained by using
Dobrushin boundary conditions.  Further, this expectation is strongly
supported by the rigorous results of Bovier and K\"{u}lske concerning the
existence for $d\ge4$ of such interface states in SOS models \cite{BK94}.
Indeed, the existence of such interface states in ferromagnets motivated
Bovier and Fr\"{o}hlich to argue that there are more spin glass pure states
in $d \ge 4$ than $d=3$ (see Sec.~6.4 of \cite{BF86}). So we need to ask
why our arguments, when applied to disordered ferromagnets, do not lead to
a contradiction for $4\le d < 8$.

In fact, they do not, for at least two reasons. One of them, already
discussed above, is the lack of appearance of interface states in the
metastate (with coupling independent boundary conditions). We have not
argued, even for spin glasses, that there are absolutely no more than two
pure states for $d < 8$, but only that not more than two appear in the
metastate (and that these are the physically relevant ones).  For both
ordinary EA spin glasses in $d < 8$ and ordinary disordered ferromagnets in
all $d$, the metastate with free or periodic (or random) b.c.'s should
exhibit no more than a single pair of pure states.  Indeed, in the highly
disordered ferromagnet even for $d >8$, the free or periodic (but not
random) b.c.~metastate is supported only on the single pair of
homogeneously magnetized ground states \cite{NS94,NS96c}.

But one might object that in the ferromagnet setting, unlike the spin glass
case, one does know how to choose b.c.'s, namely of the Dobrushin type, so
that one can obtain Gibbs states with interfaces in the thermodynamic limit
and metastates supported on them. For example, by taking a mixed b.c. with
the Dobrushin ``equator'' between the plus and minus parts of the boundary
occuring at a variety of heights with various weights, one could obtain a
metastate supported on many different pure interface states.  So, wouldn't
the fact that for $4 \le d < 8$, the zero temperature limit of the
metastate could not be supported on more than two ground states, still
contradict our reasoning that the number of ground states in the metastate
at zero temperature is an upper bound for the number of pure states at very
low temperature?

Here is where the second reason for a distinction between ferromagnets and
spin glasses comes into play.  It concerns the universality hypothesis of
our argument, that for any fixed $\lambda < \infty$, the low temperature
phase structure of a spin glass should be qualitatively the same,
regardless of the value of $\lambda$.  For spin glasses, this seems a
perfectly plausible working hypothesis; indeed, if this were not so, then,
e.g., a Gaussian distribution could have a different low $T$ thermodynamic
structure, in terms of pure state multiplicity, than a uniform distribution
on $[-J,J]$, or a $\pm J$ spin glass.  But this is not likely; it is
expected that only non-universal features such as $T_c(d)$ depend on the
coupling distribution (assuming that it is symmetric about zero and without
slowly-decaying tails).

But this is not so for disordered ferromagnets, where there are no energy
cancellations along an interface --- unlike in spin glasses.  This 
may well lead to a greater sensitivity of interface stability to the
strength of the disorder.  It seems quite plausible, as suggested to us by
Bovier and K\"{u}lske, that for some dimensions $d \ge 4$, the
usual interface states obtained through normal Dobrushin b.c.'s could
perhaps disappear for $\lambda$ above some critical value $\lambda_c <
\infty$, even for arbitrarily low temperature.  That is, for $\lambda <
\lambda_c$, the interface would be flat for both $T=0$ and small $T > 0$,
but for $\lambda > \lambda_c$, the interface would be rough for {\it any\/}
$T > 0$ (and perhaps also for $T=0$).

Our analysis suggests that this is indeed so for $d < 8$, so that there
would be only two pure states (even for the mixed Dobrushin b.c.~metastate
described above) for ordinary disordered ferromagnets at very low
temperatures, {\it providing the disorder is sufficiently (but not
infinitely) strong\/}.  For $4 \le d < 8$, this could be a consequence of
the strong disorder either completely destabilizing the interface states so
that they are entirely absent, or else of partially destabilizing them so
that finding them would require Dobrushin type b.c.'s, but with a {\it
coupling-dependent\/} equator (at a nonconstant height).

\section{Discussion}
\label{sec:discussion}

In this paper we have constructed a class of nearest-neighbor spin glass
models with (unusual) coupling distributions depending on a disorder
strength parameter $\lambda$ that itself is temperature dependent.  These
models are designed to have the property that their thermodynamic
equilibrium properties at low temperature $T$ should be the same as for
models with more familiar distributions (e.g., $\pm J$ or Gaussian), but
they have the rare advantage that their ground state structure is known.
The basic assumption of this paper is that a nearest-neighbor Ising spin
glass model with any ``reasonable'' coupling distribution --- i.e.,
symmetric about zero, and with small or zero weight on very large coupling
magnitudes --- will display qualitatively equivalent thermodynamics at
fixed $d$ (e.g., presence or absence of a phase transition, number and
organization of pure states in the spin glass phase, etc.) --- at least for
very low temperature.  While this remains an assumption, it is a common one
in theoretical spin glass studies --- so, for example, the thermodynamics
of two extreme cases, the $\pm J$ and Gaussian distributions, are usually
assumed to be the same \cite{BNE}.

All of the models discussed here have the desired properties at finite
$\lambda$ (i.e., when $T>0$).  At infinite $\lambda$ ($T=0$) they do not,
and we make no claims as to whether any {\it ground state\/} properties of
ordinary spin glasses can be inferred from these highly disordered models.
But the interesting aspect of the analysis is that, while the thermodynamic
properties at nonzero temperature cannot be directly solved for, the ground
state properties of our models can, and the analysis in
Sec.~\ref{sec:analysis} enables us to infer properties of realistic spin
glass models at low but nonzero temperatures.

In general, the number of pure states for a given system remains the same
or increases (at a phase transition point) as temperature decreases.  Of
course, this general tendency is violated at a first-order phase
transition, due to phase coexistence.  One well-known example of such a
violation is the $q$-state Potts ferromagnet with $q$ sufficiently large
(depending on $d$), which at $T_c$ has its paramagnetic pure phase
coexisting with its $q$ ordered pure phases, while below $T_c$ the
paramagnetic phase is unstable (for rigorous proofs, see
\cite{KS,M86,LMR}).  However, it remains the case, for these and other
systems with first-order transitions, that the number of pure phases above
the transition is no larger than the number below.  More interesting
re-entrant behavior occurs in other systems --- see, e.g., Ref.~\cite{ST}.
(A different type of re-entrant behavior occurs in some spin glasses in the
temperature-concentration phase diagram \cite{BY}, but this doesn't appear
to violate the general rule of number of pure states not decreasing as
temperature decreases.)

However, we are unaware of any natural examples where the number of ground
states is {\it smaller\/} than the number of pure states at arbitrarily low
temperature \cite{Shlosman}.  For the spin glass, we are looking at an even
weaker claim --- namely, that the number of pure states in the {\it
metastate\/} does not decrease as temperature is lowered.  That is, it is
sufficient to consider only incongruent pure states, as discussed in
Sec.~\ref{sec:metastate}.

These considerations enable us to draw a set of interesting conclusions
from our analysis of these models.  The most natural, and obvious, one is
that the {\it number of ground states in the zero-temperature highly
disordered metastate (e.g., with free or periodic boundary conditions)
provides an upper bound to the number of pure states at very low
temperature seen in realistic spin glass models.\/} This ground state
structure is known: the $T=0$ metastate of the highly disordered model is
supported on a single pair of ground states below eight dimensions, and
(uncountably) infinitely many above eight \cite{NS94,NS96c}.  Our analysis
therefore provides evidence in favor of the existence of no more than a
single pair of pure states at low temperatures in realistic spin glass
models below eight dimensions \cite{FS}.  (This conclusion is consistent,
of course, with there being only a single pure state, either paramagnetic
or otherwise, at all nonzero temperatures in some or all of these
dimensions.)  Our analysis does not allow us to draw conclusions about what
happens above eight dimensions.

There are two other logical possibilities, either of which would 
also be quite interesting.  It could be that our general intuition about
the behavior of the metastate as a function of temperature is violated
here, so that the number of ground states is {\it smaller\/} than the
number of pure states in some dimensions.  In other words, in these models
there might be a jump in the number of thermodynamic states, from larger to
smaller, in the metastate at zero temperature.  We cannot rule out this
possibility, other to note that the discovery of a class of models in which
this occurs raises the interesting possibility that it might occur
elsewhere also.

Indeed, a jump presumably does occur in the ordinary $2D$ EA Ising spin
glass --- but it goes in the other, more natural, direction, i.e. from
smaller to larger as temperature is lowered to zero.  The $2D$ spin glass
is believed to be paramagnetic at all nonzero temperatures, and has at
least a pair of ground states at zero temperature.  Our conclusion is that,
while we cannot logically exclude the possibility of a lowering of the
number of thermodynamic states as temperature goes to zero, it appears to
be less likely than the number remaining the same or increasing.

We mention in passing a third possibility --- that the number of
pure states in a typical spin glass metastate does not behave in a
continuous or even monotonic fashion at all as temperature is lowered
\cite{note1}.  This possibility presents a
picture of the low-temperature spin glass phase far different from any that
have appeared so far in the literature.  An extreme version of this
possibility (bearing some similarity to eigenvalue dependence on disorder
realizations in low dimensional localization), which we present primarily
for illustrative purposes, is as follows. At any (low) temperature not
depending on the coupling realization ${\cal J}$, there would be no broken
symmetry (and a unique infinite volume Gibbs state) for almost every ${\cal
J}$, but nevertheless, if a typical ${\cal J}$ were picked first and then
the temperature $T$ were varied, there would be a (countably infinite)
dense set of temperatures, {\it depending on ${\cal J}$\/}, with broken
symmetry pure phases for that ${\cal J}$ at those temperatures!  However,
because there is no evidence of such a picture to date, we do not pursue it
further here.

{\it Acknowledgments.\/} This research was supported in part by NSF Grants
DMS-98-02310 (CMN) and DMS-98-02153 (DLS).  The authors thank Anton Bovier,
Aernout van~Enter, Christof K\"{u}lske and Peter Young for useful
discussions and constructive comments.


\begin{references}
\bibitem{rigor}
A completely rigorous proof, though, has so far been obtained
only for $d=2$.
\bibitem{note1} 
This should not be confused with the milder phenomenon of chaotic
temperature dependence \cite{BM,FH86}, in which the pure states themselves
change chaotically with temperature but their number does not.
\end{references}
\end{document}